\documentclass{aastex63}
\usepackage{makecell}
\usepackage{amsmath}
\usepackage{cases}
\graphicspath{{./}{figures/}}
\begin{document}
%\begin{CJK}{UTF8}{}

%\title{Detection of the extended $\gamma$-ray emission around Calvera}
\title{Detection of the extended $\gamma$-ray emission from the high Galactic latitude Calvera's SNR candidate}

\author[0000-0001-5135-5942]{Yuliang Xin}
\author{Xiaolei Guo}
\affiliation{School of Physical Science and Technology, Southwest Jiaotong University, Chengdu 610031, China\\
\href{mailto:ylxin@swjtu.edu.cn}{ylxin@swjtu.edu.cn},
\href{mailto:xlguo@swjtu.edu.cn}{xlguo@swjtu.edu.cn}}

%\affil
%\email{ylxin@swjtu.edu.cn (YLX)}

%% Note that the \and command from previous versions of AASTeX is now
%% depreciated in this version as it is no longer necessary. AASTeX 
%% automatically takes care of all commas and "and"s between authors names.

%% AASTeX 6.3 has the new \collaboration and \nocollaboration commands to
%% provide the collaboration status of a group of authors. These commands 
%% can be used either before or after the list of corresponding authors. The
%% argument for \collaboration is the collaboration identifier. Authors are
%% encouraged to surround collaboration identifiers with ()s. The 
%% \nocollaboration command takes no argument and exists to indicate that
%% the nearby authors are not part of surrounding collaborations.

%% Mark off the abstract in the ``abstract'' environment. 
\begin{abstract}

We report the extended GeV $\gamma$-ray emission that spatially associated with the high Galactic latitude supernova remnant (SNR) candidate - Calvera's SNR with the Pass 8 data recorded by the {\em Fermi} Large Area Telescope.
The $\gamma$-ray spectrum of Calvera's SNR between 100 MeV and 1 TeV shows an evident ($\sim$ 3.4$\sigma$) spectral curvature at several tens of GeV.
The multi-wavelength data can be fitted with either a leptonic model or a hadronic one.
However, the leptonic model exhibits the inconsistent between the flat radio spectrum and the hard GeV $\gamma$-ray spectrum of Calvera's SNR.
For the hadronic model, the spectral index of protons should be harder than 1.6.
And the total energy of protons is fitted to be more than one order of magnitude higher than the explosion energy of a typical supernova, which also challenges the hadronic model.
The evident spectral curvature and the absence of non-thermal X-ray emission from Calvera's SNR makes it to be an interesting source bridging young-aged SNRs with bright non-thermal X-ray emission and old-aged SNRs interacting with molecular clouds.

\end{abstract}

%% Keywords should appear after the \end{abstract} command. 
%% See the online documentation for the full list of available subject
%% keywords and the rules for their use.
%% From the front matter, we move on to the body of the paper.
%% Sections are demarcated by \section and \subsection, respectively.
%% Observe the use of the LaTeX \label
%% command after the \subsection to give a symbolic KEY to the
%% subsection for cross-referencing in a \ref command.
%% You can use LaTeX's \ref and \label commands to keep track of
%% cross-references to sections, equations, tables, and figures.
%% That way, if you change the order of any elements, LaTeX will
%% automatically renumber them.
%%
%% We recommend that authors also use the natbib \citep
%% and \citet commands to identify citations.  The citations are
%% tied to the reference list via symbolic KEYs. The KEY corresponds
%% to the KEY in the \bibitem in the reference list below. ,，

\section{Introduction}
\label{intro}

Supernova remnants (SNRs) are widely believed to be the dominant accelerators of Galactic cosmic rays \citep[CRs;][]{1934PNAS...20..259B, 2022RvMPP...6...19L}.
Up to now, nearly three hundred of SNRs have been clearly identified mainly through the radio observations \citep{2014BASI...42...47G}. 
And among them, only ten percent is detected with $\gamma$-ray emission \citep{2019ApJ...874...50Z, 2021ApJ...910...78Z}. 
Thanks to the quick developments of the ground Cherenkov and space telescopes, especially the Large Area Telescope (LAT) on board the
{\em Fermi} satellite, more and more SNRs are detected with good spatial and spectral measurements in the $\gamma$-ray range during the past decade \citep{2014ApJ...785L..22Y,2015A&A...577A..12F,2021ApJ...908...22X,2022A&A...660A.129A}, revealing a variety of $\gamma$-ray spectra \citep{2019ApJ...874...50Z}. 
Determining the nature of the $\gamma$-ray emission plays an essential role in evaluating contributions of SNRs to the flux of Galactic cosmic rays, which requires studying individual remnants to determine both if and how they accelerate cosmic rays.
Here we report on the $\gamma$-ray detection of a recently discovered SNR associated with the high Galactic latitude Calvera pulsar.

Calvera (also known as 1RXS J141256.0+792204 or PSR J1412+7922) is a high Galactic latitude ($b$ $\sim$ 37$^{\circ}$) soft X-ray source, which was first discovered in the ROSAT All-Sky Survey and then identified to be an isolated neutron star \citep{2007Ap&SS.308..181H,2008ApJ...672.1137R}.
Using {\em XMM}-Newton data, \citet{2011MNRAS.410.2428Z} discovered X-ray pulsations with the period of {\em P} = 59.2 ms, which confirms the neutron star nature of Calvera. 
The spectral analysis shows that the X-ray emission from Calvera is thermal, which can be well reproduced by a two-component model composed of either two hydrogen atmosphere models or two blackbodies in the range of 0.1 - 0.25 keV \citep{2011MNRAS.410.2428Z}.
The following {\em Chandra} and NICER observations determined the period derivative of Calvera to be $\dot{P} = (3.19 \pm 0.08) \times 10^{-15}$ s s$^{-1}$, corresponding to the characteristic age of $\tau_{c}$ = $P/2\dot{P}$ = 294 kyr \citep{2013ApJ...778..120H,2015ApJ...812...61H,2019ApJ...877...69B}. The spin-down luminosity and the surface dipole magnetic field strength are calculated to be $\dot{E} = 6.1 \times 10^{35}$ erg s$^{-1}$ and $B_{s} = 4.4 \times 10^{11}$ G, respectively.
The deep search failed to detect the radio pulsations at the known period for Calvera, which makes it to be a radio-quiet X-ray pulsar \citep{2007A&A...476..331H,2011MNRAS.410.2428Z}.
With the {\em Fermi}-LAT data, \citet{2011MNRAS.410.2428Z} claimed to detect the $\gamma$-ray pulsations from Calvera at $>$ 100 MeV. 
However, the further analysis of additional {\em Fermi}-LAT data did not confirm the $\gamma$-ray detection, and an $\gamma$-ray upper limit was derived to be at least two orders of magnitude below the typical $\gamma$-ray luminosities of pulsars with comparable spin-down luminosity \citep{2011ApJ...736L...3H,2013ApJ...778..120H}.

Recently, a ring of low surface brightness radio emission around the Calvera pulsar was detected with the data from LOFAR Two-metre Sky Survey \citep[LoTSS;][]{2017A&A...598A.104S,2022arXiv220714141A}.
Together with the low-significance emission at 325 MHz by the Westerbork Northern Sky Survey \citep[WENSS;][]{1997A&AS..124..259R} and 1.4 GHz by the NRAO VLA Sky Survey \citep[NVSS;][]{1998AJ....115.1693C}, the spectral index of the ring is calculated to be 0.71 $\pm$ 0.09.
\citet{2022arXiv220714141A} considered three possible interpretations for the radio ring, including an HII region, a SNR, or an Odd Radio Circle (ORC). 
The flat radio spectrum makes an HII region disfavored, and the difference in size between the ring and the other known ORCs 
discards the possibility that the ring is an ORC.
Considering the positional coincidence among the ring, the Calvera pulsar and the X-ray-emitting non-equilibrium ionisation plasma around this region, the SNR interpretation is favored.
If the ring is indeed a SNR associated with Calvera pulsar, Calvera’s SNR (SNR G118.4+37.0) will be one of few SNRs in the Galactic halo. And the large heights above the Galactic plane also makes it to be an interesting source to study the SNR evolution in diffuse environments and probe the interstellar medium (ISM) of the Milky Way halo \citep{2022arXiv220714141A}. 

In the present work, we carry out a detailed analysis of the GeV $\gamma$-ray emission around the region of Calvera with the {\em Fermi}-LAT Pass 8 data. 
And this paper is organized as follows. In Section 2, we describe the data analysis routines and results, including the spatial and spectral analysis. 
A discussion about the physical origin of the $\gamma$-ray emission is presented in Section 3 based on the multi-wavelength observations, followed by the conclusion of this work in Section 4.

\section{Fermi-LAT Data Analysis}
\label{fermi}

\subsection{Data Reduction}
\label{data reduction}

We use the latest Pass 8 version of the {\em Fermi}-LAT data recorded from 2008 August 4 (Mission Elapsed Time 239557418) to 2022 March 24 (Mission Elapsed Time 669772805) to analysis the $\gamma$-ray emission around Calvera.
This analysis is performed within a $20^\circ \times 20^\circ$ square region centered at the position of Calvera pulsar.
For the spectral analysis, the energy range adopted is 100 MeV - 1 TeV, while the spatial analysis is carried out in the energy range of 1 GeV - 1 TeV in consideration of the improved LAT resolution at higher energies.
The events with zenith angles greater than $90^\circ$ are excluded to reduce the contamination from Earth Limb.
The standard {\it ScienceTools} software package \footnote {http://fermi.gsfc.nasa.gov/ssc/data/analysis/software/} with the binned analysis method, together with the instrumental response function (IRF) of ``P8R3{\_}SOURCE{\_}V3'', are adopted.
The Galactic and isotropic diffuse background emissions are modeled according to {\tt gll\_iem\_v07.fits} and {\tt iso\_P8R3\_SOURCE\_V3\_v1.txt}
\footnote {http://fermi.gsfc.nasa.gov/ssc/data/access/lat/BackgroundModels.html}, respectively.
The sources in the incremental version of the fourth {\em Fermi}-LAT source catalog \citep[4FGL-DR3;][]{2020ApJS..247...33A, 2022ApJS..260...53A}, together with the two diffuse backgrounds, are included in the model, which is generated by the user-contributed software 
{\tt make4FGLxml.py}\footnote{http://fermi.gsfc.nasa.gov/ssc/data/analysis/user/}.
For the fitting procedure, the spectral parameters and the normalizations of sources within $10^{\circ}$ around Calvera are set to be free, as well as the normalizations of the two diffuse backgrounds.

\subsection{Spatial Analysis}
\label{Spatial}

\begin{figure*}[!htb]
	\centering
	\includegraphics[width=0.48\textwidth]{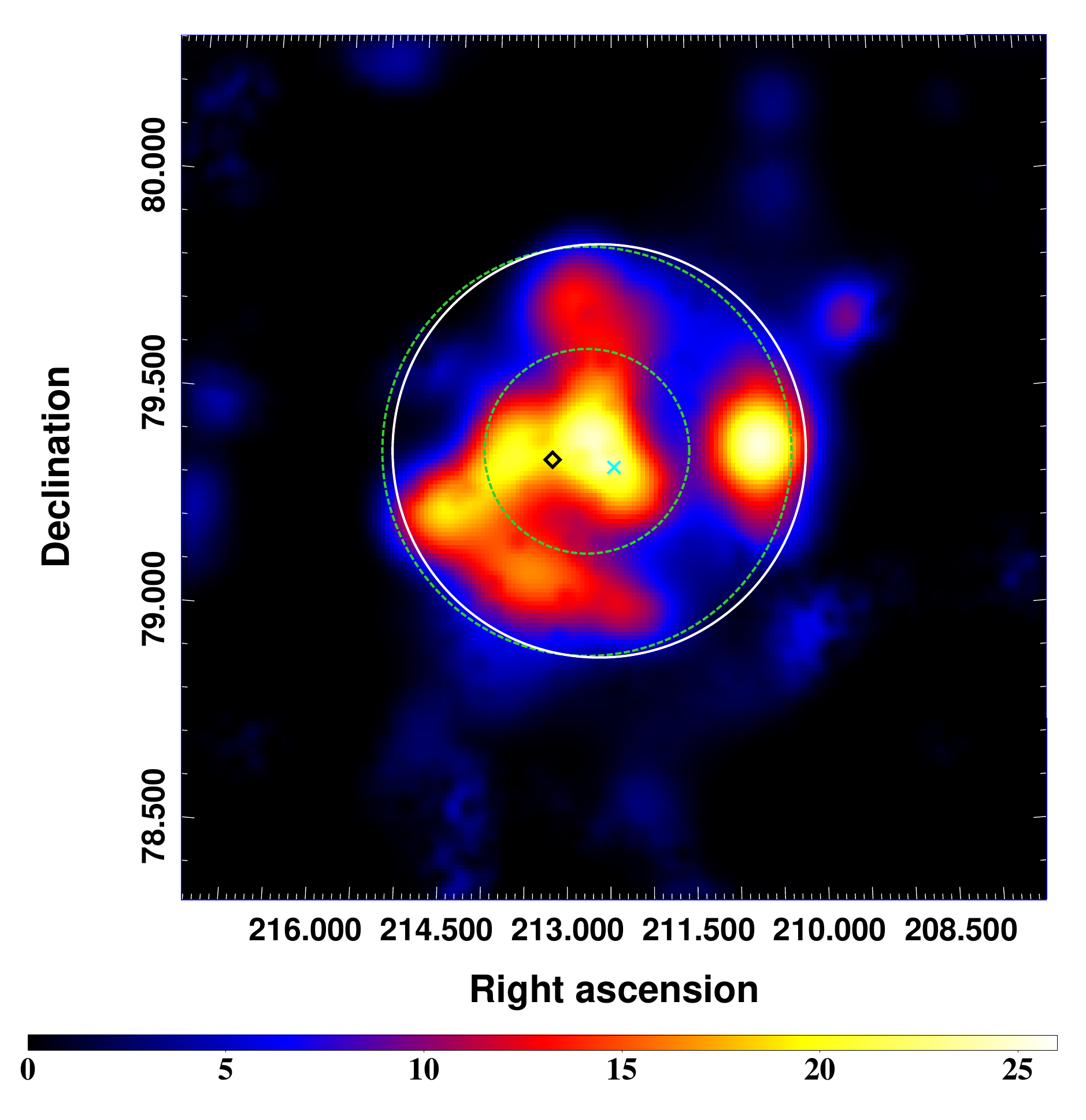}
    \includegraphics[width=0.48\textwidth]{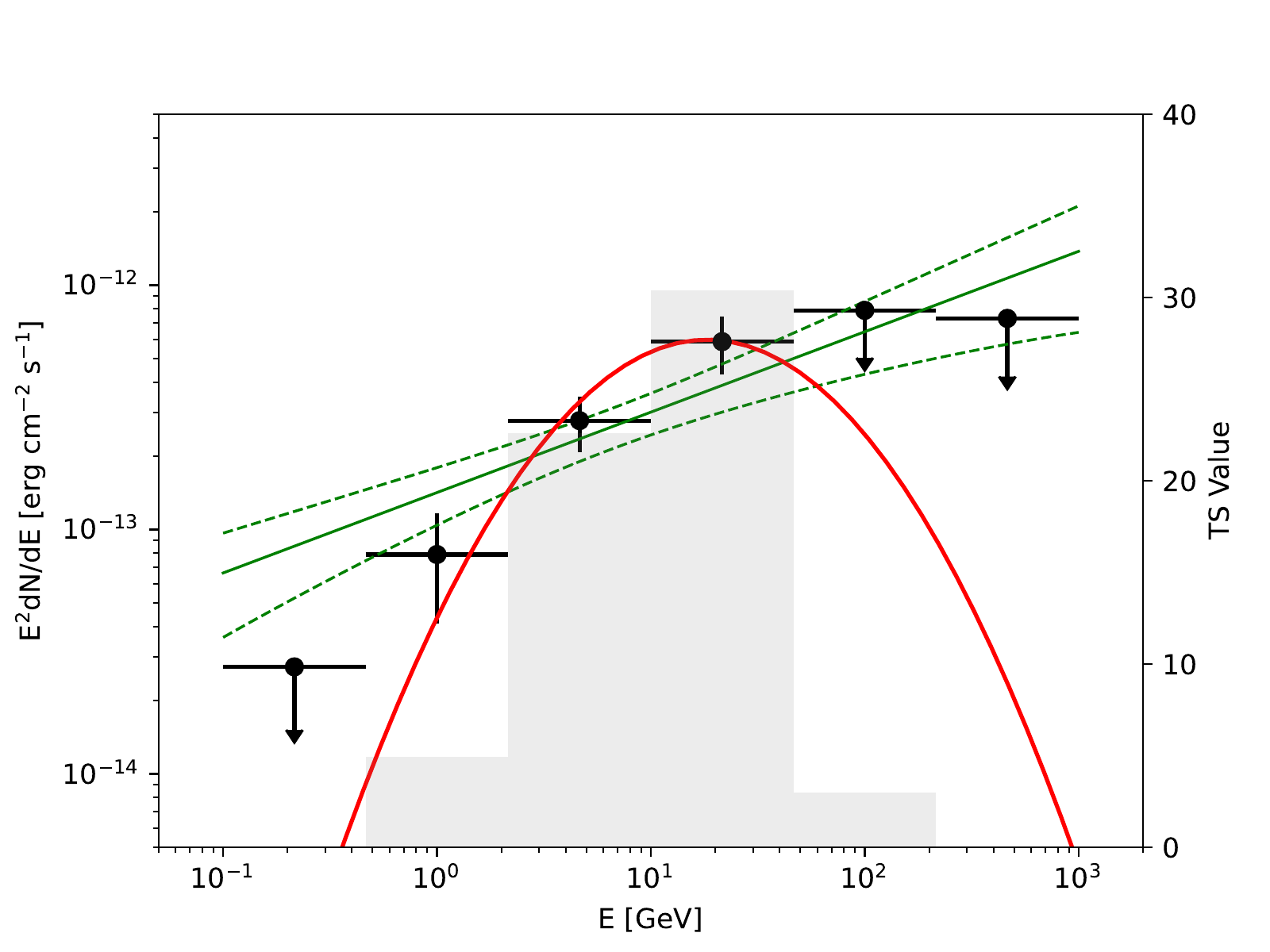}
	\caption{Left: $2^{\circ}\!.0$ $\times$ $2^{\circ}\!.0$ TSmap for photons above 1 GeV. The TSmap is smoothed with a Gaussian kernel of $\sigma$ = $0^{\circ}\!.02$. The white sold circle shows the best-fit radius of the uniform disk for the spatial template of SrcX. The black diamond reveals the Calvera pulsar PSR J1412+7922. And the cyan cross marks the fitting position of SrcX as a point source. Two green dashed circles describe the ring-morphology of the diffuse radio emission (Calvera's SNR) by LOFAR at 144 MHz \citep{2022arXiv220714141A}. Right: SED of SrcX. The black arrows indicate the 95\% upper limits and the gray histogram shows the TS value for each energy bin. The green solid and dashed lines show the best-fit power-law spectrum and its 1$\sigma$ statistic error in the energy range of 100 MeV - 1 TeV. The red sold line is the best-fit log-parabola spectrum.}
	\label{fig1:tsmap}
\end{figure*}

In the 4FGL-DR3 catalog, there is a $\gamma$-ray point source (4FGL J1409.8+7921) in the region of Calvera, which has no identified counterpart \citep{2020ApJS..247...33A, 2022ApJS..260...53A}. And the photon spectral index and integral photon flux in the energy of 1 - 100 GeV with a power law model are given to be 1.82 $\pm$ 0.22 and (4.64 $\pm$ 1.69)$\times10^{-11}$  photon cm$^{-2}$ s$^{-1}$, respectively.
With the energy range of 1 GeV - 1 TeV, we first created a $2^{\circ}\!.0$ $\times$ $2^{\circ}\!.0$ Test Statistic (TS) map with the command {\em gttsmap} by subtracting the emission from the diffuse backgrounds and all 4FGL-DR3 sources (except 4FGL J1409.8+7921) in the best-fit model, which is shown in the left panel of Figure \ref{fig1:tsmap}. 
The TS map shows significant $\gamma$-ray emission around Calvera, which is marked as SrcX afterwards. 
To test the spatial extension of the $\gamma$-ray emission from SrcX, we first adopted {\tt Fermipy}, a {\tt PYTHON} package that automates analyses with the Fermi Science Tools \citep{2017ICRC...35..824W}, to get the best-fit position of SrcX as a point source. And the corrected coordinate of SrcX is fitted to be R.A. = $212^{\circ}\!.454 \pm 0^{\circ}\!.021$, Dec. = $79^{\circ}\!.344 \pm 0^{\circ}\!.028$.
With the new position, the TS value of SrcX as a point source is 23.8.
Then we adopted the extended templates to describe the $\gamma$-ray emission from SrcX, including an uniform disk, a two-dimensional (2D) Gaussian template, and an uniform ring.
The best-fit central positions and extensions (the radius containing 68\% of the intensity; R$_{\rm 68}$) of the uniform disk and 2D-Gaussian template are given by {\tt Fermipy},
which are shown in Table \ref{table:spatial}.
For the template of the uniform ring, the central position, as well as the inner and outer radii, were adopted from the radio observation by LOFAR at 144 MHz \citep{2022arXiv220714141A}.
The TS values for the different extended templates are listed in Table \ref{table:spatial}.

We compared the overall maximum likelihood of the extended template ($\mathcal{L}_{\rm ext}$; alternative hypothesis) with that of the point source model ($\mathcal{L}_{\rm Pt}$; null hypothesis), and defined the significance of the extension as to be TS$_{\rm ext}$ = -2(log$\mathcal{L}_{\rm Pt}$ - log$\mathcal{L}_{\rm ext}$) \citep{2012ApJ...756....5L}.
The alternative hypothesis is significantly preferred to the null hypothesis only if TS$_{\rm ext}$ $>$ 16.
We found that the uniform disk centered at (R.A. = $212^{\circ}\!.649$, Dec. = $79^{\circ}\!.389$) with a radius of $0^{\circ}\!.478$ can best fit the $\gamma$-ray emission from SrcX. And the improved significance is TS$_{\rm ext}$ = 30.0, which corresponds to $\sim$5.5 $\sigma$ extension with one additional degree of freedom relative to the point source model.
With the extended template of uniform disk, the TS value of SrcX is fitted to be 54.1 in the energy range of 1 GeV - 1 TeV, corresponding to a significance level of $\sim$6.4 $\sigma$ with five degrees of freedom.

\begin{table}[!htb]
\centering
\caption {Spatial Analysis for SrcX between 1 GeV and 1 TeV}
\begin{tabular}{ccccccc}
\hline \hline
Spatial Template    &  R.A., Dec.    & TS Value    & Degrees of Freedom   & -log(Likelihood)\\
%                    & & R$_{\rm 68}$
% &                               &$\rm 10^{-11}$ ph/$\rm cm^{2}$/s  &  Value &  Freedom   \\
\hline
Point Source      & $212^{\circ}\!.454 \pm 0^{\circ}\!.021$, $79^{\circ}\!.344 \pm 0^{\circ}\!.028$  &  23.8  & 4   & 122016.00 \\ 
\hline
Uniform disk  & \makecell[c]{$212^{\circ}\!.649 \pm 0^{\circ}\!.038$, $79^{\circ}\!.389 \pm 0^{\circ}\!.038$, \\ R$_{\rm 68}$ = $0^{\circ}\!.392 ^{+0^{\circ}\!.025}_{-0^{\circ}\!.025}$}  & 54.1  & 5  & 122001.07 \\
\hline
2D-Gaussian   & \makecell[c]{$212^{\circ}\!.361 \pm 0^{\circ}\!.071$, $79^{\circ}\!.358 \pm 0^{\circ}\!.068$, \\ R$_{\rm 68}$ = $0^{\circ}\!.394 ^{+0^{\circ}\!.068}_{-0^{\circ}\!.050}$}  & 49.2  & 5  & 122003.51\\
\hline
Uniform Ring    & \makecell[c]{$212^{\circ}\!.803$, $79^{\circ}\!.388$,\\ R$_{\rm inner}$ = $14^{\prime}\!.2$, R$_{\rm outer}$ = $28^{\prime}\!.4$}  & 46.1  & 6  & 122005.03 \\
\hline
\end{tabular}
\label{table:spatial}
\end{table}   

\subsection{Spectral Analysis}
\label{Spectral}

We performed a spectral analysis of SrcX in the energy range from 100 MeV to 1 TeV with the spatial template of the uniform disk.
With a single power-law (PL; $dN/dE \propto E^{-\alpha}$) spectrum, the TS value of SrcX is calculated to be $\rm TS_{PL} =$ 44.4. And the spectral index and integrated photon flux of SrcX are fitted to be 1.67 $\pm$ 0.10 and $(6.16 \pm 2.10)\times10^{-10}$ 
photon cm$^{-2}$ s$^{-1}$, respectively.
We then test for the spectral curvature in the spectrum of SrcX using a log-parabola (LogPb; $dN/dE \propto E^{-(\alpha+\beta {\rm log}(E/E_b))}$) spectrum. And the TS value of SrcX is fitted to be $\rm TS_{LogPb} =$ 55.8.
The variation of TS values with the LogPb model is $\rm \Delta TS$ = $\rm TS_{LogPb}$ - $\rm TS_{PL}$ = 11.4, which suggests an evidence of spectral curvature for the $\gamma$-ray emission of SrcX ($\sim $3.4$\sigma$ improvement with respect to a PL model).
The fitting of LogPb model presents the spectral parameters of $\alpha = 2.05 \pm 0.27$ and $\beta = 0.31 \pm 0.14$. And the integrated photon flux is calculated to be $(1.45 \pm 0.52)\times10^{-10}$ 
photon cm$^{-2}$ s$^{-1}$, respectively.

Furthermore, to obtain the spectral energy distribution (SED) of SrcX, the data from 100 MeV to 1 TeV are divided into 6 logarithmic equal energy bins, and the same likelihood fitting analysis was performed for each energy bin.
For the energy bin with TS value of SrcX smaller than 5.0, an upper limit with 95\% confidence level is calculated.
The results of the SED are shown in the right panel of Figure \ref{fig1:tsmap}, and the global fitting with the PL and LogPb models are also overplotted.

\section{Discussion}
\label{discussion}

\begin{figure*}[!htb]
	\centering
        \includegraphics[width=0.6\textwidth]{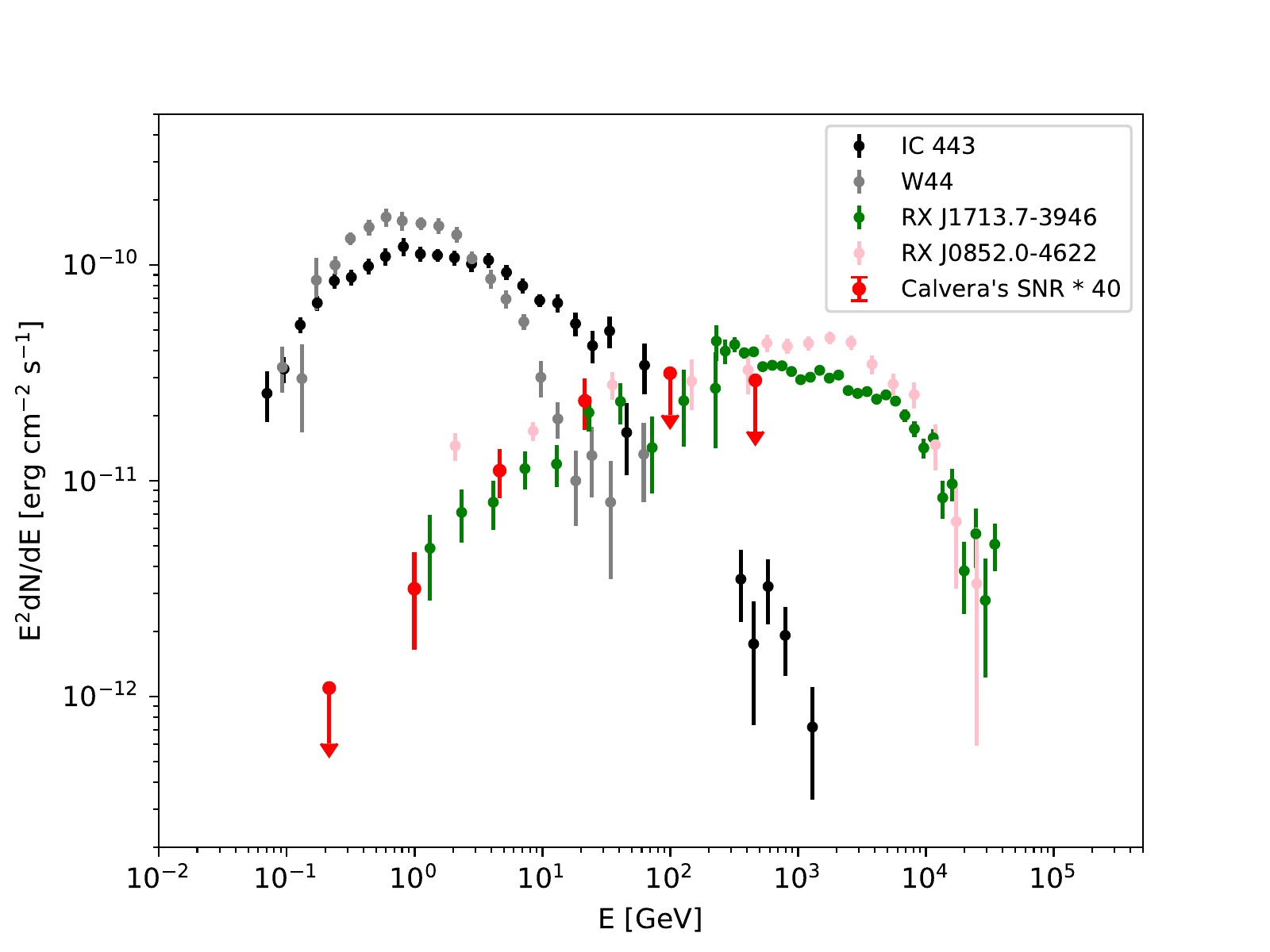}
	\caption{Typical $\gamma$-ray SEDs of several prominent SNRs. The energy fluxes of Calvera's SNR are scaled upward by 40 times for comparison.}
	\label{fig2:sed-multiSNRs}
\end{figure*}

The above data analysis shows that the extension of the $\gamma$-ray emission from SrcX is well consistent with the radio size of the Calvera's SNR, as shown in the left panel of Figure \ref{fig1:tsmap}, which supports SrcX as the GeV counterpart of Calvera's SNR.
Figure \ref{fig2:sed-multiSNRs} shows the $\gamma$-ray spectra of several {\em Fermi}-LAT observed SNRs, together with Calvera's SNR.
The old-aged SNRs interacting with molecular clouds, like IC 443 and W44 \citep{2013Sci...339..807A}, show the spectral break at $\sim$ GeV, whose $\gamma$-ray emission are suggested to be produced by the decay of neutral pions due to the inelastic {\em pp} collisions (hadronic model).
Another class of SNRs with hard GeV $\gamma$-ray spectra, including RX J1713.7-3946 \citep{2018A&A...612A...6H} and RX J0852.0-4622 \citep{2018A&A...612A...7H}, shows the spectral curvature at $\sim$ TeV. These SNRs are typically young-aged systems with strong non-thermal X-ray emission, and the $\gamma$-ray emission from them are suggested to be from inverse Compton scattering (ICS) of accelerated electrons (leptonic model).
The GeV $\gamma$-ray spectrum of Calvera's SNR shown in Figure \ref{fig2:sed-multiSNRs} is similar to that of the young-aged SNRs. 
However, the absence of the non-thermal X-ray emission makes it different from them.
Moreover, the spectral analysis in Section \ref{Spectral} shows an evident spectral curvature at tens of GeV for the $\gamma$-ray spectrum of Calvera's SNR, which is much lower than that of these young-aged SNRs with the spectral curvature at $\sim$ TeV.
Therefore, Calvera's SNR may be one class of SNRs bridging young-aged SNRs with bright non-thermal X-ray emission and old-aged SNRs interacting with molecular clouds.

\begin{figure*}[!htb]
	\centering
	\includegraphics[width=0.48\textwidth]{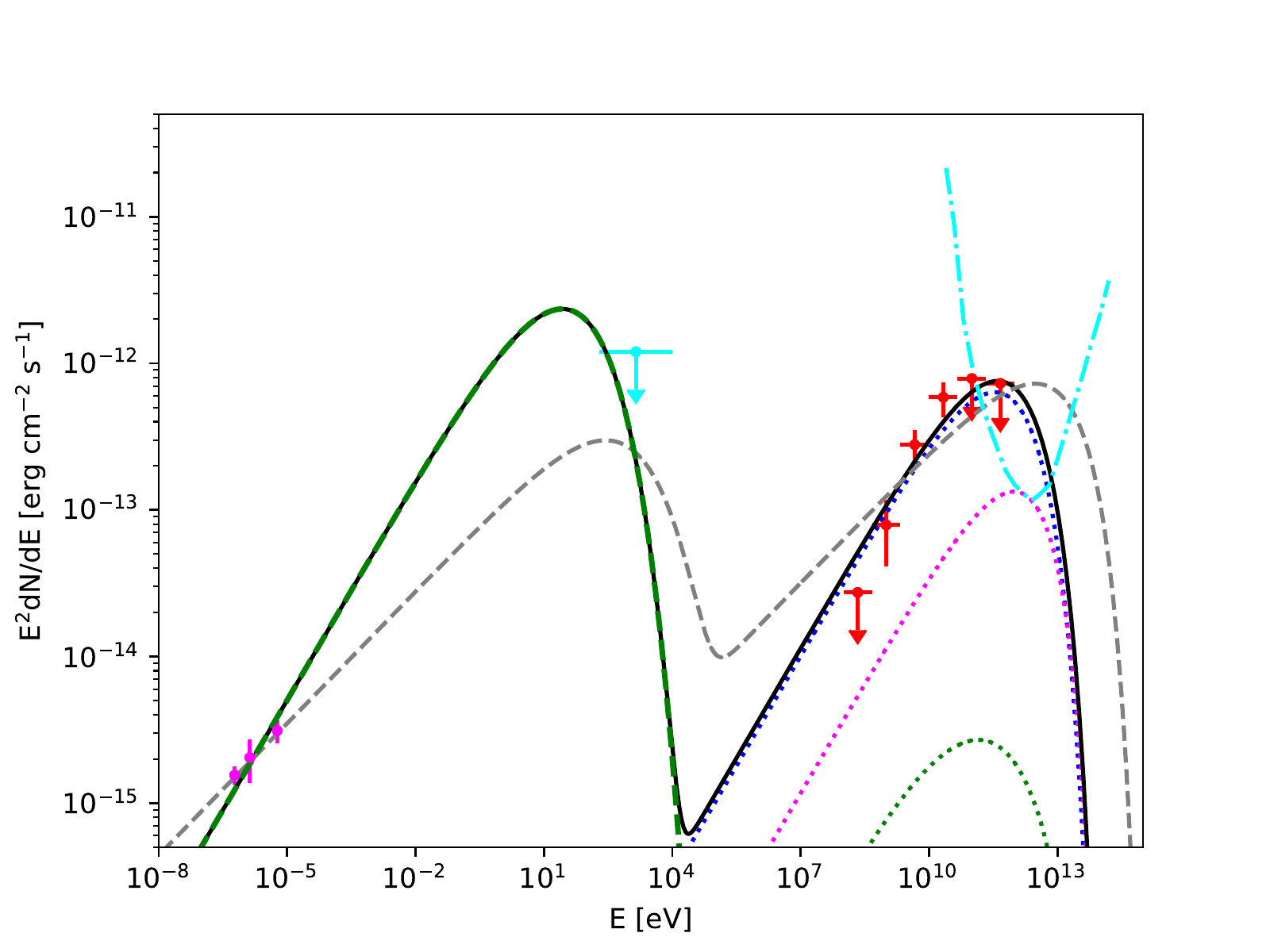}
        \includegraphics[width=0.48\textwidth]{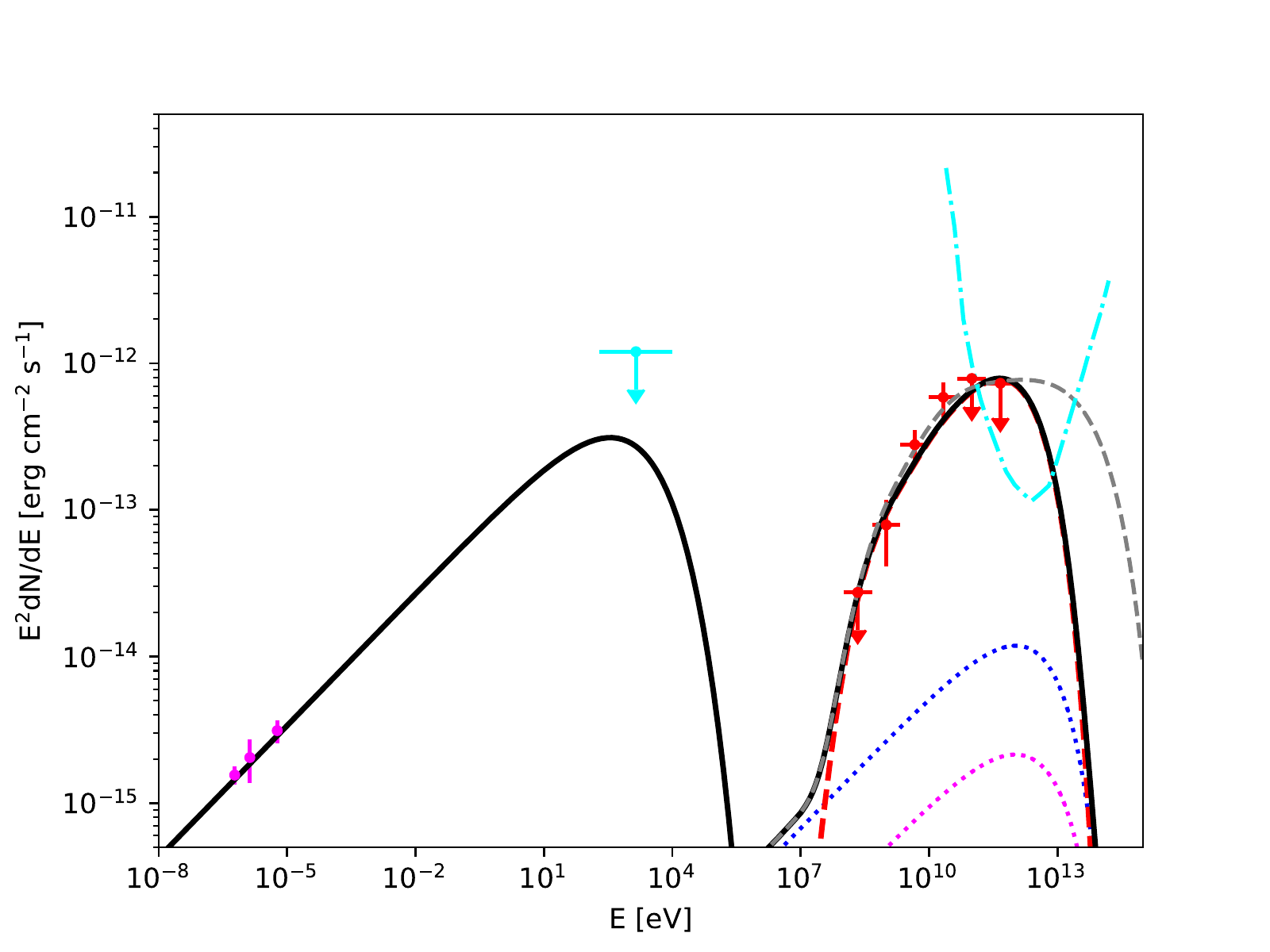}
	\caption{Modeling of the multi-wavelength SED of Calvera's SNR. The left panel is for the leptonic model, where the power law spectra with indices of 2.0 and 2.4 for electrons are presented by the black solid and gray dashed lines, respectively. The radio emission is dominated by the synchrotron component, shown as the green dashed curve. For $\gamma$-ray emission, the contribution of the ICS emission includes three components from CMB (blue dashed), infrared (magenta dashed), and optical (green dashed) radiation fields. The right panel is for the hadronic model, where the PL and BPL models are presented by the black solid and gray dashed lines, respectively. The contribution from pion-decay process with PL model is shown as the red dashed line. The flux of bremsstrahlung component is too low to show, and the black solid line represents the sum of the different radiation components. The gray dot-dashed line shows the differential sensitivity of CTA-North \citep[50 hr;][]{2019scta.book.....C}.}
	\label{fig3:model}
\end{figure*}

To explore the origin of the $\gamma$-ray emission from Calvera's SNR, both hadronic and leptonic models are considered here. 
And for the leptonic model, the $\gamma$-ray emission from the bremsstrahlung process of relativistic electrons is also calculated, together with the ICS component.
The radio and non-thermal X-ray emission are produced by high-energy electrons via the synchrotron process.
And for ICS process, three components of the interstellar radiation field are taken into account including the cosmic microwave background (CMB), the infrared and optical blackbody components. The temperature and energy density of the infrared and optical blackbody components are (T$_1$ = 30 K, u$_1$ = 0.1 eV cm$^{-3}$) and (T$_2$ = 5000 K, u$_2$ = 0.1 eV cm$^{-3}$), respectively, considering the large heights above the Galactic plane for SrcX \citep{2006ApJ...648L..29P, 2008ApJ...682..400P}.
The components of the different radiation mechanisms are computed using the {\em naima} package \citep{2015ICRC...34..922Z}.

The distance of Calvera's SNR is adopted to be 3.3 kpc, which is estimated by assuming a neutron star radius of 13 km, and fitting the thermal emission from Calvera pulsar \citep{2021ApJ...922..253M}.
With this distance, the physical radius of SrcX is estimated to be about 27.5 pc.
For the SNRs in the Galactic halo, the ambient densities are typical very low.
Using the model of \citet{2013ApJ...770..118M} with the distance between Calvera and the center of the Galaxy of r = 9.5 kpc,
\citet{2022arXiv220714141A} estimated the ambient density in the halo around Calvera to be n$_{\rm gas}$ = $4 \times 10^{-4}$ cm$^{-3}$. Here, we adopt this value to calculate the  hadronic and bremsstrahlung components.

The radio data of Calvera's SNR are from \citet{2022arXiv220714141A}, including the observations by LoTSS at 144 MHz, WENSS at 325 MHz and NVSS at 1.4 GHz.
Considering the absence of the non-thermal X-ray emission from Calvera's SNR, we adopt the thermal X-ray emission from Calvera pulsar with $\rm 1.20 \times 10^{-12}$ erg cm$^{-2}$ s$^{-1}$ in the energy range of 0.2 - 10 keV as an upper limit to constrain the model parameters \citep{2021ApJ...922..253M}.

For both leptonic and hadronic models, the spectra of electrons and protons are first assumed to be a single power-law with an exponential cutoff (PL) in the form of 
\begin{equation}
    \frac{dN_{\rm i}}{dE} \propto \left(\frac{E}{E_0}\right)^{-\alpha_{i}} exp\left(-\frac{E}{E_{\rm i, cut}}\right) 
\end{equation}
where $\alpha_{\rm i}$ and $\rm E_{i, cut}$ are the spectral index and the cutoff energy of particles, respectively, for i = e or p.
The radio spectral index of 0.71 $\pm$ 0.09 would suggest the spectral index of electrons to be $\sim$ 2.4 for the synchrotron process. And it also corresponds to a $\gamma$-ray spectral index of 1.7 for ICS component, and 2.4 for hadronic component assuming the same index for electrons and protons.
For the leptonic model, the spectral index of electrons with 2.4 is first adopted to fit the multi-wavelength data with other parameters listed in Table \ref{table:model}. And the fitting SED is shown as the gray dashed line in the left panel of Figure \ref{fig3:model}.
The fitting result suggests a harder distribution for electrons considering the $\gamma$-ray spectrum, especially the upper limit in the low energy range.
Then we changed the spectral index of electrons to be 2.0 considering the statistic uncertainties of the radio data.
And the magnetic field strength of 6 $\mu$G is needed to explain the radio flux of Calvera's SNR.
Considering the upper limits in the X-ray and $\gamma$-ray bands, the cutoff energy of electrons needs to be less than $\sim$ 10 TeV. And the value should be much lower if the spectral curvature at tens of GeV of the $\gamma$-ray spectrum is taken into account.
With the value of B = 6 $\mu$G and E$_{\rm e, cut}$ = 10 TeV, the synchrotron cooling timescale of electrons is estimated to be about 3.4 $\times$ 10$^{\rm 4}$ yrs, which is much larger than the age of $\sim$ 1400 yr for Calvera's SNR assuming the Sedov-Taylor phase evolution and the ambient density of $4 \times 10^{-4}$ cm$^{-3}$ \citep{2022arXiv220714141A}. 
Here the age is recalculated with the same evolution model and a radius of 27 pc instead of 54 pc used by \citet{2022arXiv220714141A}.
Moreover, it should be noted that the age estimation depends on the ambient density around SNR and higher density value would result in an older age for Calvera's SNR.
The total energy of electrons above 1 GeV is calculated to be W$_{\rm e}$ = 1.8 $\times$ 10$^{\rm 47}$ erg. And such value is similar to that of other young-aged SNRs, which are of the order of 10$^{\rm 47}$ - 10$^{\rm 48}$ erg \citep{2014A&A...567A..23Y}.

Considering the evident curvature of the $\gamma$-ray spectrum likely produced by the synchrotron cooling of electrons, the model with electron distribution to be a broken power law (BPL) spectrum is also taken into account, with the form of
\begin{equation}
\frac{dN_{\rm i}}{dE} \propto exp\left(-\frac{E}{E_{\rm i,cut}}\right) % \left\{
\begin{cases}
\left(\frac{E}{E_0}\right)^{-\alpha_{i1}} \qquad \qquad \qquad \qquad;  E < E_{\rm i,break}  \\
\left(\frac{E_{\rm i,break}}{E_0}\right)^{\alpha_{i2}-\alpha_{i1}} \left(\frac{E}{E_0}\right)^{-\alpha_{i2}} \,\,\,\,; E \geq E_{\rm i,break}
\end{cases}
\end{equation}
However, the break energy, E$_{\rm e,break}$, is calculated to be about 248 TeV with a magnetic field strength of 6 $\mu$G, by assuming the synchrotron cooling timescale equal to the age of Calvera's SNR of 1400 yr. And this is conflict with the upper limits in the X-ray and $\gamma$-ray bands.

The fitting multi-wavelength SED with hadronic model is compiled in the right panel of Figure \ref{fig3:model}. 
In this model, the spectral index of electrons is set to be 2.4, which is consistent with the flat radio spectrum of Calvera's SNR.
The cutoff energy of electrons E$_{\rm e, cut}$ is calculated by equalling the synchrotron cooling timescale and the age of Calvera's SNR to reduce the number of free parameters, which is given by
E$_{\rm e, cut}$ = 1.25 $\times$ 10$^{\rm 7}$ t$_{\rm age; yr}^{-1}$ B$_{\rm \mu G}^{\rm -2}$ TeV. 
The radio flux and the upper limit in the X-ray band constrain the total energy of electrons above 1 GeV and the magnetic field strength to be W$_{\rm e}$ = 1.6 $\times$ 10$^{\rm 46}$ erg and B $\simeq$ 15 $\mu$G, respectively. 
And the corresponding cutoff energy of electrons is calculated to be 39.7 TeV.
To explain the hard GeV $\gamma$-ray spectrum of Calvera's SNR, the spectral index of protons is fitted to be about 1.6.
The upper limits in hundreds of GeV constrains the cutoff energy of protons to be less than $\sim$ 20 TeV.
Due to the very low  ambient density in the $\gamma$-ray region, the total energy of protons above 1 GeV, W$_{\rm p}$, is estimated to be about 4.6 $\times$ 10$^{\rm 52}$ (n$_{\rm gas}$/4 $\times$ 10$^{-4}$ cm$^{\rm -3}$)$^{-1}$ erg, which is more than one order of magnitude higher than the explosion energy of a typical supernova ($\sim$ $10^{51}$ erg). 
It should be noted that the uncertainty of the distance measured with the thermal emission comes from the whole neutron star surface would result in the variation of the Galactic height for Calvera's SNR \citep{2021ApJ...922..253M}. A lower Galactic height with higher ambient density could be expected. And the much higher total energy of protons in the hadronic model may attribute to the underestimate of the ambient density around Calvera's SNR.

Similar to the leptonic model, a broken power law spectrum for protons is also tested considering the expectation of a break may produced by the escape of protons at an advanced age of SNR \citep{2010A&A...513A..17O}.
The fitting SED is shown as the gray dashed line in the right panel of Figure \ref{fig3:model}, with the parameters listed in Table \ref{table:model}.
The spectral indices below and above the break are adopted to be $\alpha_{\rm p1}$ = 1.5 and $\alpha_{\rm p2}$ = 2.0, with the break energy is fitted to be about 0.5 TeV. The cutoff energy of protons with 1 PeV can not be well constrained, and the future possible detection in the TeV band by Cherenkov Telescope Array in the northern hemisphere \citep[CTA-North;][]{2019scta.book.....C} would be helpful.

\begin{table*}
	\centering
	%\tabletypesize{\small}
	\caption {Parameters for the models}
	\begin{tabular}{cccccccccc}
		\hline \hline
		Model & $\alpha_e$ & $\alpha_p$ & E$_{p,\rm break}$ & E$_{e,\rm cut}$ & E$_{p, \rm cut}$ & W$_e$  & W$_p$ & $B$ \\
		      &  &   & (TeV)        & (TeV)      & (TeV)        & ($10^{47}$ erg) & ($10^{52}$ erg) & ($\mu$G) \\
		\hline
		Leptonic  & $2.4$  & $-$ & $-$  & $100$ & $-$  & $6.3$ & $-$ & $1.8$ \\
		 (PL)     & $2.0$  & $-$ & $-$  & $10$  & $-$  & $1.8$ & $-$ & $6.0$ \\
		\hline
		Hadronic(PL)  & $2.4$ & $1.6$     & $-$    & $39.7$  & $20$   & $0.16$  & $4.6$ & $15.0$ \\
		Hadronic(BPL) & $2.4$ & $1.5/2.0$ & $0.5$  & $39.7$  & $1000$ & $0.16$  & $6.4$ & $15.0$ \\
		\hline
		\hline
	\end{tabular}
	\label{table:model}
	\tablecomments{The total energy of relativistic particles, $W_{e,p}$, is 
		calculated for $E > 1$ GeV.}
\end{table*}

%\subsection{PWN scenario}
Except the SNR scenario for the $\gamma$-ray emission from SrcX, we also consider the possibility that the emission comes from a pulsar wind nebula (PWN) powered by Calvera pulsar, although there is no diffuse radio or X-ray emission detected around Calvera pulsar.
The $\gamma$-ray PWNe detected by {\em Fermi}-LAT are typically driven by the energetic pulsars with spin-down power between 10$^{36}$ and 10$^{39}$ erg s$^{-1}$ \citep{2013ApJ...773...77A}.
However, the spin-down luminosity of Calvera pulsar is $\dot{E} = 6.1 \times 10^{35}$ erg s$^{-1}$, which is slightly lower than that.
%In addition, the $\gamma$-ray emission from the typical PWNe are usually center bright, which is not consistent with the $\gamma$-ray morphology of SrcX.
In addition, the $\gamma$-ray luminosity of SrcX in the energy range of 10 - 316 GeV is calculated to be 1.75 $\times$ 10$^{33}$ erg s$^{-1}$, which is much lower than that of the typical $\gamma$-ray PWNe with above 10$^{34}$ erg s$^{-1}$ \citep{2013ApJ...773...77A}.
%All evidences above suggest that the $\gamma$-ray emission from SrcX disfavors the PWN origin.
Nonetheless, the hypothetical PWN may also have partial contribution to the $\gamma$-ray emission from SrcX, considering the $\gamma$-ray morphology with spot emission around Calvera pulsar, as shown in the left panel of Figure \ref{fig1:tsmap}. And the future observations searching for the possible PWN in the radio/X-ray band may help to clear the $\gamma$-ray origin from SrcX.

\section{Conclusions}
\label{con}

In this work, we report the detection of the $\gamma$-ray emission in the direction of Calvera's SNR with more than 13 yr of Pass 8 data recorded by the {\em Fermi}-LAT. The $\gamma$-ray spatial morphology of Calvera's SNR is significant extended ($\sim$6.4 $\sigma$). And it can be well described by an uniform disk with a radius of $0^{\circ}\!.478$, which is well consistent with the radio size of the remnant.
The $\gamma$-ray spectrum of Calvera's SNR follows a hard power-law form with an index of 1.67 $\pm$ 0.10 in the energy range of 100 MeV - 1 TeV. The higher TS value with a log-parabola form suggests an evidence spectral curvature for the $\gamma$-ray emission from Calvera's SNR.
The hard $\gamma$-ray spectrum of Calvera's SNR makes it similar to the typical young-aged SNRs, whose $\gamma$-ray emission comes from the ICS process.
However, there is no non-thermal X-ray emission detected around Calvera's SNR, which is different from these SNRs.
Considering the spectral curvature at tens of GeV, Calvera's SNR may be one class of SNRs bridging young-aged SNRs with bright non-thermal X-ray emission and old-aged SNRs interacting with molecular clouds.
Both leptonic and hadronic models could marginally explain the multi-wavelength data of Calvera's SNR.
However, the flat radio spectrum seems to be not well consistent with the hard GeV $\gamma$-ray spectrum in the leptonic model.
And for the hadronic model, the spectral index of protons should be harder than 1.6, which is difficult to produce in the conventional diffusive shock acceleration model of strong shocks. 
Moreover, Calvera's SNR is located in the high Galactic latitude, which makes the ambient density around it very low.
And the total energy of protons above 1 GeV is calculated to be about W$_{\rm p}$ = 4.6 $\times$ 10$^{\rm 52}$ (n$_{\rm gas}$/4 $\times$ 10$^{-4}$ cm$^{\rm -3}$)$^{-1}$ erg, which is more than one order of magnitude higher than the explosion energy of a typical supernova ($\sim$ $10^{51}$ erg). 
The ratio of the normalization of electron distribution to that of protons, $K_{\rm ep}$, is much lower than the locally measured cosmic ray electron to proton flux ratio ($\sim$ 0.01).

Calvera's SNR provides a good target to study the SNR evolution and particle acceleration in the high Galactic latitude, and probe the interstellar medium of the Milky Way halo. 
More multi-wavelength observations in future, especially the possible non-thermal X-ray emission and very-high-energy $\gamma$-ray emission by CTA-North, are crucial to ultimately understanding its nature.

\acknowledgments

%We would like to thank the anonymous referee for very helpful comments, which help to improve the paper. 
This work is supported by the Natural Science Foundation for Young Scholars of Sichuan Province, China (No. 2022NSFSC1808),
the Science and Technology Department of Sichuan Province (No. 2021YFSY0031, No.2020YFSY0016),
the Fundamental Research Funds for the Central Universities (No. 2682021CX073, No. 2682021CX074, No. 2682022ZTPY013), 
and the National Natural Science Foundation of China under the grants 12103040 and 12147208.

%% To help institutions obtain information on the effectiveness of their 
%% telescopes the AAS Journals has created a group of keywords for telescope 
%% facilities.
%
%% Following the acknowledgments section, use the following syntax and the
%% \facility{} or \facilities{} macros to list the keywords of facilities used 
%% in the research for the paper.  Each keyword is check against the master 
%% list during copy editing.  Individual instruments can be provided in 
%% parentheses, after the keyword, but they are not verified.

%% Appendix material should be preceded with a single \appendix command.
%% There should be a \section command for each appendix. Mark appendix
%% subsections with the same markup you use in the main body of the paper.

%% Each Appendix (indicated with \section) will be lettered A, B, C, etc.
%% The equation counter will reset when it encounters the \appendix
%% command and will number appendix equations (A1), (A2), etc. The
%% Figure and Table counter will not reset.

\bibliography{sample63}{}

\begin{thebibliography}{}
\expandafter\ifx\csname natexlab\endcsname\relax\def\natexlab#1{#1}\fi
\providecommand{\url}[1]{\href{#1}{#1}}
\providecommand{\dodoi}[1]{doi:~\href{http://doi.org/#1}{\nolinkurl{#1}}}
\providecommand{\doeprint}[1]{\href{http://ascl.net/#1}{\nolinkurl{http://ascl.net/#1}}}
\providecommand{\doarXiv}[1]{\href{https://arxiv.org/abs/#1}{\nolinkurl{https://arxiv.org/abs/#1}}}

\bibitem[{{Abdollahi} {et~al.}(2020){Abdollahi}, {Acero}, {Ackermann},
  {Ajello}, {Atwood}, {Axelsson}, {Baldini}, {Ballet}, {Barbiellini},
  {Bastieri}, {Becerra Gonzalez}, {Bellazzini}, {Berretta}, {Bissaldi},
  {Blandford}, {Bloom}, {Bonino}, {Bottacini}, {Brandt}, {Bregeon}, {Bruel},
  {Buehler}, {Burnett}, {Buson}, {Cameron}, {Caputo}, {Caraveo}, {Casandjian},
  {Castro}, {Cavazzuti}, {Charles}, {Chaty}, {Chen}, {Cheung}, {Chiaro},
  {Ciprini}, {Cohen-Tanugi}, {Cominsky}, {Coronado-Bl{\'a}zquez}, {Costantin},
  {Cuoco}, {Cutini}, {D'Ammando}, {DeKlotz}, {de la Torre Luque}, {de Palma},
  {Desai}, {Digel}, {Di Lalla}, {Di Mauro}, {Di Venere}, {Dom{\'\i}nguez},
  {Dumora}, {Fana Dirirsa}, {Fegan}, {Ferrara}, {Franckowiak}, {Fukazawa},
  {Funk}, {Fusco}, {Gargano}, {Gasparrini}, {Giglietto}, {Giommi}, {Giordano},
  {Giroletti}, {Glanzman}, {Green}, {Grenier}, {Griffin}, {Grondin}, {Grove},
  {Guiriec}, {Harding}, {Hayashi}, {Hays}, {Hewitt}, {Horan},
  {J{\'o}hannesson}, {Johnson}, {Kamae}, {Kerr}, {Kocevski}, {Kovac'evic'},
  {Kuss}, {Landriu}, {Larsson}, {Latronico}, {Lemoine-Goumard}, {Li},
  {Liodakis}, {Longo}, {Loparco}, {Lott}, {Lovellette}, {Lubrano}, {Madejski},
  {Maldera}, {Malyshev}, {Manfreda}, {Marchesini}, {Marcotulli},
  {Mart{\'\i}-Devesa}, {Martin}, {Massaro}, {Mazziotta}, {McEnery}, {Mereu},
  {Meyer}, {Michelson}, {Mirabal}, {Mizuno}, {Monzani}, {Morselli},
  {Moskalenko}, {Negro}, {Nuss}, {Ojha}, {Omodei}, {Orienti}, {Orlando},
  {Ormes}, {Palatiello}, {Paliya}, {Paneque}, {Pei}, {Pe{\~n}a-Herazo},
  {Perkins}, {Persic}, {Pesce-Rollins}, {Petrosian}, {Petrov}, {Piron}, {Poon},
  {Porter}, {Principe}, {Rain{\`o}}, {Rando}, {Razzano}, {Razzaque}, {Reimer},
  {Reimer}, {Remy}, {Reposeur}, {Romani}, {Saz Parkinson}, {Schinzel},
  {Serini}, {Sgr{\`o}}, {Siskind}, {Smith}, {Spandre}, {Spinelli}, {Strong},
  {Suson}, {Tajima}, {Takahashi}, {Tak}, {Thayer}, {Thompson}, {Tibaldo},
  {Torres}, {Torresi}, {Valverde}, {Van Klaveren}, {van Zyl}, {Wood},
  {Yassine}, \& {Zaharijas}}]{2020ApJS..247...33A}
{Abdollahi}, S., {Acero}, F., {Ackermann}, M., {et~al.} 2020, \apjs, 247, 33,
  \dodoi{10.3847/1538-4365/ab6bcb}

\bibitem[{{Abdollahi} {et~al.}(2022){Abdollahi}, {Acero}, {Baldini}, {Ballet},
  {Bastieri}, {Bellazzini}, {Berenji}, {Berretta}, {Bissaldi}, {Blandford},
  {Bloom}, {Bonino}, {Brill}, {Britto}, {Bruel}, {Burnett}, {Buson}, {Cameron},
  {Caputo}, {Caraveo}, {Castro}, {Chaty}, {Cheung}, {Chiaro}, {Cibrario},
  {Ciprini}, {Coronado-Bl{\'a}zquez}, {Crnogorcevic}, {Cutini}, {D'Ammando},
  {De Gaetano}, {Digel}, {Di Lalla}, {Dirirsa}, {Di Venere}, {Dom{\'\i}nguez},
  {Fallah Ramazani}, {Fegan}, {Ferrara}, {Fiori}, {Fleischhack}, {Franckowiak},
  {Fukazawa}, {Funk}, {Fusco}, {Galanti}, {Gammaldi}, {Gargano}, {Garrappa},
  {Gasparrini}, {Giacchino}, {Giglietto}, {Giordano}, {Giroletti}, {Glanzman},
  {Green}, {Grenier}, {Grondin}, {Guillemot}, {Guiriec}, {Gustafsson},
  {Harding}, {Hays}, {Hewitt}, {Horan}, {Hou}, {J{\'o}hannesson}, {Karwin},
  {Kayanoki}, {Kerr}, {Kuss}, {Landriu}, {Larsson}, {Latronico},
  {Lemoine-Goumard}, {Li}, {Liodakis}, {Longo}, {Loparco}, {Lott}, {Lubrano},
  {Maldera}, {Malyshev}, {Manfreda}, {Mart{\'\i}-Devesa}, {Mazziotta}, {Mereu},
  {Meyer}, {Michelson}, {Mirabal}, {Mitthumsiri}, {Mizuno}, {Moiseev},
  {Monzani}, {Morselli}, {Moskalenko}, {Negro}, {Nuss}, {Omodei}, {Orienti},
  {Orlando}, {Paneque}, {Pei}, {Perkins}, {Persic}, {Pesce-Rollins},
  {Petrosian}, {Pillera}, {Poon}, {Porter}, {Principe}, {Rain{\`o}}, {Rando},
  {Rani}, {Razzano}, {Razzaque}, {Reimer}, {Reimer}, {Reposeur},
  {S{\'a}nchez-Conde}, {Saz Parkinson}, {Scotton}, {Serini}, {Sgr{\`o}},
  {Siskind}, {Smith}, {Spandre}, {Spinelli}, {Sueoka}, {Suson}, {Tajima},
  {Tak}, {Thayer}, {Thompson}, {Torres}, {Troja}, {Valverde}, {Wood}, \&
  {Zaharijas}}]{2022ApJS..260...53A}
{Abdollahi}, S., {Acero}, F., {Baldini}, L., {et~al.} 2022, \apjs, 260, 53,
  \dodoi{10.3847/1538-4365/ac6751}

\bibitem[{{Acero} {et~al.}(2022){Acero}, {Lemoine-Goumard}, \&
  {Ballet}}]{2022A&A...660A.129A}
{Acero}, F., {Lemoine-Goumard}, M., \& {Ballet}, J. 2022, \aap, 660, A129,
  \dodoi{10.1051/0004-6361/202142200}

\bibitem[{{Acero} {et~al.}(2013){Acero}, {Ackermann}, {Ajello}, {Allafort},
  {Baldini}, {Ballet}, {Barbiellini}, {Bastieri}, {Bechtol}, {Bellazzini},
  {Blandford}, {Bloom}, {Bonamente}, {Bottacini}, {Brandt}, {Bregeon},
  {Brigida}, {Bruel}, {Buehler}, {Buson}, {Caliandro}, {Cameron}, {Caraveo},
  {Cecchi}, {Charles}, {Chaves}, {Chekhtman}, {Chiang}, {Chiaro}, {Ciprini},
  {Claus}, {Cohen-Tanugi}, {Conrad}, {Cutini}, {Dalton}, {D'Ammando}, {de
  Palma}, {Dermer}, {Di Venere}, {Silva}, {Drell}, {Drlica-Wagner}, {Falletti},
  {Favuzzi}, {Fegan}, {Ferrara}, {Focke}, {Franckowiak}, {Fukazawa}, {Funk},
  {Fusco}, {Gargano}, {Gasparrini}, {Giglietto}, {Giordano}, {Giroletti},
  {Glanzman}, {Godfrey}, {Gr{\'e}goire}, {Grenier}, {Grondin}, {Grove},
  {Guiriec}, {Hadasch}, {Hanabata}, {Harding}, {Hayashida}, {Hayashi}, {Hays},
  {Hewitt}, {Hill}, {Horan}, {Hou}, {Hughes}, {Inoue}, {Jackson}, {Jogler},
  {J{\'o}hannesson}, {Johnson}, {Kamae}, {Kawano}, {Kerr}, {Kn{\"o}dlseder},
  {Kuss}, {Lande}, {Larsson}, {Latronico}, {Lemoine-Goumard}, {Longo},
  {Loparco}, {Lovellette}, {Lubrano}, {Marelli}, {Massaro}, {Mayer},
  {Mazziotta}, {McEnery}, {Mehault}, {Michelson}, {Mitthumsiri}, {Mizuno},
  {Monte}, {Monzani}, {Morselli}, {Moskalenko}, {Murgia}, {Nakamori}, {Nemmen},
  {Nuss}, {Ohsugi}, {Okumura}, {Orienti}, {Orlando}, {Ormes}, {Paneque},
  {Panetta}, {Perkins}, {Pesce-Rollins}, {Piron}, {Pivato}, {Porter},
  {Rain{\`o}}, {Rando}, {Razzano}, {Reimer}, {Reimer}, {Reposeur}, {Ritz},
  {Roth}, {Rousseau}, {Saz Parkinson}, {Schulz}, {Sgr{\`o}}, {Siskind},
  {Smith}, {Spandre}, {Spinelli}, {Suson}, {Takahashi}, {Takeuchi}, {Thayer},
  {Thayer}, {Thompson}, {Tibaldo}, {Tibolla}, {Tinivella}, {Torres}, {Tosti},
  {Troja}, {Uchiyama}, {Vandenbroucke}, {Vasileiou}, {Vianello}, {Vitale},
  {Werner}, {Winer}, {Wood}, \& {Yang}}]{2013ApJ...773...77A}
{Acero}, F., {Ackermann}, M., {Ajello}, M., {et~al.} 2013, \apj, 773, 77,
  \dodoi{10.1088/0004-637X/773/1/77}

\bibitem[{{Ackermann} {et~al.}(2013){Ackermann}, {Ajello}, {Allafort},
  {Baldini}, {Ballet}, {Barbiellini}, {Baring}, {Bastieri}, {Bechtol},
  {Bellazzini}, {Blandford}, {Bloom}, {Bonamente}, {Borgland}, {Bottacini},
  {Brandt}, {Bregeon}, {Brigida}, {Bruel}, {Buehler}, {Busetto}, {Buson},
  {Caliandro}, {Cameron}, {Caraveo}, {Casandjian}, {Cecchi}, {{\c{C}}elik},
  {Charles}, {Chaty}, {Chaves}, {Chekhtman}, {Cheung}, {Chiang}, {Chiaro},
  {Cillis}, {Ciprini}, {Claus}, {Cohen-Tanugi}, {Cominsky}, {Conrad}, {Corbel},
  {Cutini}, {D'Ammando}, {de Angelis}, {de Palma}, {Dermer}, {do Couto e
  Silva}, {Drell}, {Drlica-Wagner}, {Falletti}, {Favuzzi}, {Ferrara},
  {Franckowiak}, {Fukazawa}, {Funk}, {Fusco}, {Gargano}, {Germani},
  {Giglietto}, {Giommi}, {Giordano}, {Giroletti}, {Glanzman}, {Godfrey},
  {Grenier}, {Grondin}, {Grove}, {Guiriec}, {Hadasch}, {Hanabata}, {Harding},
  {Hayashida}, {Hayashi}, {Hays}, {Hewitt}, {Hill}, {Hughes}, {Jackson},
  {Jogler}, {J{\'o}hannesson}, {Johnson}, {Kamae}, {Kataoka}, {Katsuta},
  {Kn{\"o}dlseder}, {Kuss}, {Lande}, {Larsson}, {Latronico}, {Lemoine-Goumard},
  {Longo}, {Loparco}, {Lovellette}, {Lubrano}, {Madejski}, {Massaro}, {Mayer},
  {Mazziotta}, {McEnery}, {Mehault}, {Michelson}, {Mignani}, {Mitthumsiri},
  {Mizuno}, {Moiseev}, {Monzani}, {Morselli}, {Moskalenko}, {Murgia},
  {Nakamori}, {Nemmen}, {Nuss}, {Ohno}, {Ohsugi}, {Omodei}, {Orienti},
  {Orlando}, {Ormes}, {Paneque}, {Perkins}, {Pesce-Rollins}, {Piron}, {Pivato},
  {Rain{\`o}}, {Rando}, {Razzano}, {Razzaque}, {Reimer}, {Reimer}, {Ritz},
  {Romoli}, {S{\'a}nchez-Conde}, {Schulz}, {Sgr{\`o}}, {Simeon}, {Siskind},
  {Smith}, {Spandre}, {Spinelli}, {Stecker}, {Strong}, {Suson}, {Tajima},
  {Takahashi}, {Takahashi}, {Tanaka}, {Thayer}, {Thayer}, {Thompson},
  {Thorsett}, {Tibaldo}, {Tibolla}, {Tinivella}, {Troja}, {Uchiyama}, {Usher},
  {Vandenbroucke}, {Vasileiou}, {Vianello}, {Vitale}, {Waite}, {Werner},
  {Winer}, {Wood}, {Wood}, {Yamazaki}, {Yang}, \&
  {Zimmer}}]{2013Sci...339..807A}
{Ackermann}, M., {Ajello}, M., {Allafort}, A., {et~al.} 2013, Science, 339,
  807, \dodoi{10.1126/science.1231160}

\bibitem[{{Arias} {et~al.}(2022){Arias}, {Botteon}, {Bassa}, {van der Jagt},
  {van Weeren}, {O'Sullivan}, {Bosschaart}, {Dullaart}, {Hardcastle},
  {Hessels}, {Shimwell}, {Slob}, {Sturm}, {Tasse}, {Theijssen}, \&
  {Vink}}]{2022arXiv220714141A}
{Arias}, M., {Botteon}, A., {Bassa}, C.~G., {et~al.} 2022, arXiv e-prints,
  arXiv:2207.14141.
\newblock \doarXiv{2207.14141}

\bibitem[{{Baade} \& {Zwicky}(1934)}]{1934PNAS...20..259B}
{Baade}, W., \& {Zwicky}, F. 1934, Proceedings of the National Academy of
  Science, 20, 259, \dodoi{10.1073/pnas.20.5.259}

\bibitem[{{Bogdanov} {et~al.}(2019){Bogdanov}, {Ho}, {Enoto}, {Guillot},
  {Harding}, {Jaisawal}, {Malacaria}, {Manthripragada}, {Arzoumanian}, \&
  {Gendreau}}]{2019ApJ...877...69B}
{Bogdanov}, S., {Ho}, W. C.~G., {Enoto}, T., {et~al.} 2019, \apj, 877, 69,
  \dodoi{10.3847/1538-4357/ab1b2e}

\bibitem[{{Cherenkov Telescope Array Consortium} {et~al.}(2019){Cherenkov
  Telescope Array Consortium}, {Acharya}, {Agudo}, {Al Samarai}, {Alfaro},
  {Alfaro}, {Alispach}, {Alves Batista}, {Amans}, {Amato}, {Ambrosi},
  {Antolini}, {Antonelli}, {Aramo}, {Araya}, {Armstrong}, {Arqueros},
  {Arrabito}, {Asano}, {Ashley}, {Backes}, {Balazs}, {Balbo}, {Ballester},
  {Ballet}, {Bamba}, {Barkov}, {Barres de Almeida}, {Barrio}, {Bastieri},
  {Becherini}, {Belfiore}, {Benbow}, {Berge}, {Bernardini}, {Bernardini},
  {Bernardos}, {Bernl{\"o}hr}, {Bertucci}, {Biasuzzi}, {Bigongiari}, {Biland},
  {Bissaldi}, {Biteau}, {Blanch}, {Blazek}, {Boisson}, {Bolmont}, {Bonanno},
  {Bonardi}, {Bonavolont{\`a}}, {Bonnoli}, {Bosnjak}, {B{\"o}ttcher},
  {Braiding}, {Bregeon}, {Brill}, {Brown}, {Brun}, {Brunetti}, {Buanes},
  {Buckley}, {Bugaev}, {B{\"u}hler}, {Bulgarelli}, {Bulik}, {Burton},
  {Burtovoi}, {Busetto}, {Canestrari}, {Capalbi}, {Capitanio}, {Caproni},
  {Caraveo}, {C{\'a}rdenas}, {Carlile}, {Carosi}, {Carqu{\'\i}n}, {Carr},
  {Casanova}, {Cascone}, {Catalani}, {Catalano}, {Cauz}, {Cerruti}, {Chadwick},
  {Chaty}, {Chaves}, {Chen}, {Chen}, {Chernyakova}, {Chikawa}, {Christov},
  {Chudoba}, {Cie{\'s}lar}, {Coco}, {Colafrancesco}, {Colin}, {Conforti},
  {Connaughton}, {Conrad}, {Contreras}, {Cortina}, {Costa}, {Costantini},
  {Cotter}, {Covino}, {Crocker}, {Cuadra}, {Cuevas}, {Cumani}, {D'A{\`\i}},
  {D'Ammando}, {D'Avanzo}, {D'Urso}, {Daniel}, {Davids}, {Dawson}, {Dazzi}, {De
  Angelis}, {de C{\'a}ssia dos Anjos}, {De Cesare}, {De Franco}, {de Gouveia
  Dal Pino}, {de la Calle}, {de los Reyes Lopez}, {De Lotto}, {De Luca}, {De
  Lucia}, {de Naurois}, {de O{\~n}a Wilhelmi}, {De Palma}, {De Persio}, {de
  Souza}, {Deil}, {Del Santo}, {Delgado}, {della Volpe}, {Di Girolamo}, {Di
  Pierro}, {Di Venere}, {D{\'\i}az}, {Dib}, {Diebold}, {Djannati-Ata{\"\i}},
  {Dom{\'\i}nguez}, {Dominis Prester}, {Dorner}, {Doro}, {Drass}, {Dravins},
  {Dubus}, {Dwarkadas}, {Ebr}, {Eckner}, {Egberts}, {Einecke}, {Ekoume},
  {Els{\"a}sser}, {Ernenwein}, {Espinoza}, {Evoli}, {Fairbairn},
  {Falceta-Goncalves}, {Falcone}, {Farnier}, {Fasola}, {Fedorova}, {Fegan},
  {Fernandez-Alonso}, {Fern{\'a}ndez-Barral}, {Ferrand}, {Fesquet},
  {Filipovic}, {Fioretti}, {Fontaine}, {Fornasa}, {Fortson}, {Freixas
  Coromina}, {Fruck}, {Fujita}, {Fukazawa}, {Funk}, {F{\"u}{\ss}ling},
  {Gabici}, {Gadola}, {Gallant}, {Garcia}, {Garcia L{\'o}pez}, {Garczarczyk},
  {Gaskins}, {Gasparetto}, {Gaug}, {Gerard}, {Giavitto}, {Giglietto}, {Giommi},
  {Giordano}, {Giro}, {Giroletti}, {Giuliani}, {Glicenstein}, {Gnatyk},
  {Godinovic}, {Goldoni}, {G{\'o}mez-Vargas}, {Gonz{\'a}lez}, {Gonz{\'a}lez},
  {G{\"o}tz}, {Graham}, {Grandi}, {Granot}, {Green}, {Greenshaw}, {Griffiths},
  {Gunji}, {Hadasch}, {Hara}, {Hardcastle}, {Hassan}, {Hayashi}, {Hayashida},
  {Heller}, {Helo}, {Hermann}, {Hinton}, {Hnatyk}, {Hofmann}, {Holder},
  {Horan}, {H{\"o}randel}, {Horns}, {Horvath}, {Hovatta}, {Hrabovsky},
  {Hrupec}, {Humensky}, {H{\"u}tten}, {Iarlori}, {Inada}, {Inome}, {Inoue},
  {Inoue}, {Inoue}, {Iocco}, {Ioka}, {Iori}, {Ishio}, {Iwamura}, {Jamrozy},
  {Janecek}, {Jankowsky}, {Jean}, {Jung-Richardt}, {Jurysek}, {Kaaret},
  {Karkar}, {Katagiri}, {Katz}, {Kawanaka}, {Kazanas}, {Kh{\'e}lifi}, {Kieda},
  {Kimeswenger}, {Kimura}, {Kisaka}, {Knapp}, {Kn{\"o}dlseder}, {Koch},
  {Kohri}, {Komin}, {Kosack}, {Kraus}, {Krause}, {Krau{\ss}}, {Kubo}, {Kukec
  Mezek}, {Kuroda}, {Kushida}, {La Palombara}, {Lamanna}, {Lang}, {Lapington},
  {Le Blanc}, {Leach}, {Lees}, {Lefaucheur}, {Leigui de Oliveira}, {Lenain},
  {Lico}, {Limon}, {Lindfors}, {Lohse}, {Lombardi}, {Longo}, {L{\'o}pez},
  {L{\'o}pez-Coto}, {Lu}, {Lucarelli}, {Luque-Escamilla}, {Lyard}, {Maccarone},
  {Maier}, {Majumdar}, {Malaguti}, {Mandat}, {Maneva}, {Manganaro}, {Mangano},
  {Marcowith}, {Mar{\'\i}n}, {Markoff}, {Mart{\'\i}}, {Martin},
  {Mart{\'\i}nez}, {Mart{\'\i}nez}, {Masetti}, {Masuda}, {Maurin}, {Maxted},
  {Mazin}, {Medina}, {Melandri}, {Mereghetti}, {Meyer}, {Minaya}, {Mirabal},
  {Mirzoyan}, {Mitchell}, {Mizuno}, {Moderski}, {Mohammed}, {Mohrmann},
  {Montaruli}, {Moralejo}, {Morcuende-Parrilla}, {Mori}, {Morlino}, {Morris},
  {Morselli}, {Moulin}, {Mukherjee}, {Mundell}, {Murach}, {Muraishi}, {Murase},
  {Nagai}, {Nagataki}, {Nagayoshi}, {Naito}, {Nakamori}, {Nakamura}, {Niemiec},
  {Nieto}, {Niko{\l}ajuk}, {Nishijima}, {Noda}, {Nosek}, {Novosyadlyj},
  {Nozaki}, {O'Brien}, {Oakes}, {Ohira}, {Ohishi}, {Ohm}, {Okazaki}, {Okumura},
  {Ong}, {Orienti}, {Orito}, {Osborne}, {Ostrowski}, {Otte}, {Oya}, {Padovani},
  {Paizis}, {Palatiello}, {Palatka}, {Paoletti}, {Paredes}, {Pareschi},
  {Parsons}, {Pe'er}, {Pech}, {Pedaletti}, {Perri}, {Persic}, {Petrashyk},
  {Petrucci}, {Petruk}, {Peyaud}, {Pfeifer}, {Piano}, {Pisarski}, {Pita},
  {Pohl}, {Polo}, {Pozo}, {Prandini}, {Prast}, {Principe}, {Prokhorov},
  {Prokoph}, {Prouza}, {P{\"u}hlhofer}, {Punch}, {P{\"u}rckhauer}, {Queiroz},
  {Quirrenbach}, {Rain{\`o}}, {Razzaque}, {Reimer}, {Reimer}, {Reisenegger},
  {Renaud}, {Rezaeian}, {Rhode}, {Ribeiro}, {Rib{\'o}}, {Richtler}, {Rico},
  {Rieger}, {Riquelme}, {Rivoire}, {Rizi}, {Rodriguez}, {Rodriguez Fernandez},
  {Rodr{\'\i}guez V{\'a}zquez}, {Rojas}, {Romano}, {Romeo}, {Rosado}, {Rovero},
  {Rowell}, {Rudak}, {Rugliancich}, {Rulten}, {Sadeh}, {Safi-Harb}, {Saito},
  {Sakaki}, {Sakurai}, {Salina}, {S{\'a}nchez-Conde}, {Sandaker}, {Sandoval},
  {Sangiorgi}, {Sanguillon}, {Sano}, {Santander}, {Sarkar}, {Satalecka},
  {Saturni}, {Schioppa}, {Schlenstedt}, {Schneider}, {Schoorlemmer},
  {Schovanek}, {Schulz}, {Schussler}, {Schwanke}, {Sciacca}, {Scuderi},
  {Seitenzahl}, {Semikoz}, {Sergijenko}, {Servillat}, {Shalchi}, {Shellard},
  {Sidoli}, {Siejkowski}, {Sillanp{\"a}{\"a}}, {Sironi}, {Sitarek}, {Sliusar},
  {Slowikowska}, {Sol}, {Stamerra}, {Stani{\v{c}}}, {Starling}, {Stawarz},
  {Stefanik}, {Stephan}, {Stolarczyk}, {Stratta}, {Straumann}, {Suomijarvi},
  {Supanitsky}, {Tagliaferri}, {Tajima}, {Tavani}, {Tavecchio}, {Tavernet},
  {Tayabaly}, {Tejedor}, {Temnikov}, {Terada}, {Terrier}, {Terzic}, {Teshima},
  {Testa}, {Thoudam}, {Tian}, {Tibaldo}, {Tluczykont}, {Todero Peixoto},
  {Tokanai}, {Tomastik}, {Tonev}, {Tornikoski}, {Torres}, {Torresi}, {Tosti},
  {Tothill}, {Tovmassian}, {Travnicek}, {Trichard}, {Trifoglio}, {Troyano
  Pujadas}, {Tsujimoto}, {Umana}, {Vagelli}, {Vagnetti}, {Valentino},
  {Vallania}, {Valore}, {van Eldik}, {Vandenbroucke}, {Varner}, {Vasileiadis},
  {Vassiliev}, {V{\'a}zquez Acosta}, {Vecchi}, {Vega}, {Vercellone}, {Veres},
  {Vergani}, {Verzi}, {Vettolani}, {Viana}, {Vigorito}, {Villanueva}, {Voelk},
  {Vollhardt}, {Vorobiov}, {Vrastil}, {Vuillaume}, {Wagner}, {Wagner},
  {Walter}, {Ward}, {Warren}, {Watson}, {Werner}, {White}, {White},
  {Wierzcholska}, {Wilcox}, {Will}, {Williams}, {Wischnewski}, {Wood},
  {Yamamoto}, {Yamazaki}, {Yanagita}, {Yang}, {Yoshida}, {Yoshiike},
  {Yoshikoshi}, {Zacharias}, {Zaharijas}, {Zampieri}, {Zandanel}, {Zanin},
  {Zavrtanik}, {Zavrtanik}, {Zdziarski}, {Zech}, {Zechlin}, {Zhdanov},
  {Ziegler}, \& {Zorn}}]{2019scta.book.....C}
{Cherenkov Telescope Array Consortium}, {Acharya}, B.~S., {Agudo}, I., {et~al.}
  2019, {Science with the Cherenkov Telescope Array}, \dodoi{10.1142/10986}

\bibitem[{{Condon} {et~al.}(1998){Condon}, {Cotton}, {Greisen}, {Yin},
  {Perley}, {Taylor}, \& {Broderick}}]{1998AJ....115.1693C}
{Condon}, J.~J., {Cotton}, W.~D., {Greisen}, E.~W., {et~al.} 1998, \aj, 115,
  1693, \dodoi{10.1086/300337}

\bibitem[{{Federici} {et~al.}(2015){Federici}, {Pohl}, {Telezhinsky},
  {Wilhelm}, \& {Dwarkadas}}]{2015A&A...577A..12F}
{Federici}, S., {Pohl}, M., {Telezhinsky}, I., {Wilhelm}, A., \& {Dwarkadas},
  V.~V. 2015, \aap, 577, A12, \dodoi{10.1051/0004-6361/201424947}

\bibitem[{{Green}(2014)}]{2014BASI...42...47G}
{Green}, D.~A. 2014, Bulletin of the Astronomical Society of India, 42, 47.
\newblock \doarXiv{1409.0637}

\bibitem[{{H.~E.~S.~S. Collaboration} {et~al.}(2018{\natexlab{a}}){H.~E.~S.~S.
  Collaboration}, {Abdalla}, {Abramowski}, {Aharonian}, {Ait Benkhali},
  {Akhperjanian}, {Andersson}, {Ang{\"u}ner}, {Arrieta}, {Aubert}, \&
  et~al.}]{2018A&A...612A...6H}
{H.~E.~S.~S. Collaboration}, {Abdalla}, H., {Abramowski}, A., {et~al.}
  2018{\natexlab{a}}, \aap, 612, A6, \dodoi{10.1051/0004-6361/201629790}

\bibitem[{{H.~E.~S.~S. Collaboration} {et~al.}(2018{\natexlab{b}}){H.~E.~S.~S.
  Collaboration}, {Abdalla}, {Abramowski}, {Aharonian}, {Ait Benkhali},
  {Akhperjanian}, {Ang{\"u}ner}, {Arakawa}, {Arrieta}, {Aubert}, {Backes},
  {Balzer}, {Barnard}, {Becherini}, {Becker Tjus}, {Berge}, {Bernhard},
  {Bernl{\"o}hr}, {Blackwell}, {B{\"o}ttcher}, {Boisson}, {Bolmont}, {Bordas},
  {Bregeon}, {Brun}, {Brun}, {Bryan}, {B{\"u}chele}, {Bulik}, {Capasso},
  {Carr}, {Casanova}, {Cerruti}, {Chakraborty}, {Chalme-Calvet}, {Chaves},
  {Chen}, {Chevalier}, {Chr{\'e}tien}, {Coffaro}, {Colafrancesco}, {Cologna},
  {Condon}, {Conrad}, {Cui}, {Davids}, {Decock}, {Degrange}, {Deil}, {Devin},
  {deWilt}, {Dirson}, {Djannati-Ata{\"\i}}, {Domainko}, {Donath}, {Drury},
  {Dutson}, {Dyks}, {Edwards}, {Egberts}, {Eger}, {Ernenwein}, {Eschbach},
  {Farnier}, {Fegan}, {Fernandes}, {Fiasson}, {Fontaine}, {F{\"o}rster},
  {Funk}, {F{\"u}{\ss}ling}, {Gabici}, {Gajdus}, {Gallant}, {Garrigoux},
  {Giavitto}, {Giebels}, {Glicenstein}, {Gottschall}, {Goyal}, {Grondin},
  {Hahn}, {Haupt}, {Hawkes}, {Heinzelmann}, {Henri}, {Hermann}, {Hervet},
  {Hinton}, {Hofmann}, {Hoischen}, {Holler}, {Horns}, {Ivascenko}, {Iwasaki},
  {Jacholkowska}, {Jamrozy}, {Janiak}, {Jankowsky}, {Jankowsky}, {Jingo},
  {Jogler}, {Jouvin}, {Jung-Richardt}, {Kastendieck}, {Katarzy{\'n}ski},
  {Katsuragawa}, {Katz}, {Kerszberg}, {Khangulyan}, {Kh{\'e}lifi}, {Kieffer},
  {King}, {Klepser}, {Klochkov}, {Klu{\'z}niak}, {Kolitzus}, {Komin}, {Krakau},
  {Kraus}, {Kr{\"u}ger}, {Laffon}, {Lamanna}, {Lau}, {Lees}, {Lefaucheur},
  {Lefranc}, {Lemi{\`e}re}, {Lemoine-Goumard}, {Lenain}, {Leser}, {Lohse},
  {Lorentz}, {Liu}, {L{\'o}pez-Coto}, {Lypova}, {Marandon}, {Marcowith},
  {Mariaud}, {Marx}, {Maurin}, {Maxted}, {Mayer}, {Meintjes}, {Meyer},
  {Mitchell}, {Moderski}, {Mohamed}, {Mohrmann}, {Mor{\r{a}}}, {Moulin},
  {Murach}, {Nakashima}, {de Naurois}, {Niederwanger}, {Niemiec}, {Oakes},
  {O'Brien}, {Odaka}, {{\"O}ttl}, {Ohm}, {Ostrowski}, {Oya}, {Padovani},
  {Panter}, {Parsons}, {Paz Arribas}, {Pekeur}, {Pelletier}, {Perennes},
  {Petrucci}, {Peyaud}, {Piel}, {Pita}, {Poon}, {Prokhorov}, {Prokoph},
  {P{\"u}hlhofer}, {Punch}, {Quirrenbach}, {Raab}, {Reimer}, {Reimer},
  {Renaud}, {de los Reyes}, {Richter}, {Rieger}, {Romoli}, {Rowell}, {Rudak},
  {Rulten}, {Sahakian}, {Saito}, {Salek}, {Sanchez}, {Santangelo}, {Sasaki},
  {Schlickeiser}, {Sch{\"u}ssler}, {Schulz}, {Schwanke}, {Schwemmer},
  {Seglar-Arroyo}, {Settimo}, {Seyffert}, {Shafi}, {Shilon}, {Simoni}, {Sol},
  {Spanier}, {Spengler}, {Spies}, {Stawarz}, {Steenkamp}, {Stegmann}, {Stycz},
  {Sushch}, {Takahashi}, {Tavernet}, {Tavernier}, {Taylor}, {Terrier},
  {Tibaldo}, {Tiziani}, {Tluczykont}, {Trichard}, {Tsuji}, {Tuffs}, {Uchiyama},
  {van der Walt}, {van Eldik}, {van Rensburg}, {van Soelen}, {Vasileiadis},
  {Veh}, {Venter}, {Viana}, {Vincent}, {Vink}, {Voisin}, {V{\"o}lk},
  {Vuillaume}, {Wadiasingh}, {Wagner}, {Wagner}, {Wagner}, {White},
  {Wierzcholska}, {Willmann}, {W{\"o}rnlein}, {Wouters}, {Yang}, {Zabalza},
  {Zaborov}, {Zacharias}, {Zanin}, {Zdziarski}, {Zech}, {Zefi}, {Ziegler}, \&
  {{\.Z}ywucka}}]{2018A&A...612A...7H}
---. 2018{\natexlab{b}}, \aap, 612, A7, \dodoi{10.1051/0004-6361/201630002}

\bibitem[{{Haberl}(2007)}]{2007Ap&SS.308..181H}
{Haberl}, F. 2007, \apss, 308, 181, \dodoi{10.1007/s10509-007-9342-x}

\bibitem[{{Halpern}(2011)}]{2011ApJ...736L...3H}
{Halpern}, J.~P. 2011, \apjl, 736, L3, \dodoi{10.1088/2041-8205/736/1/L3}

\bibitem[{{Halpern} {et~al.}(2013){Halpern}, {Bogdanov}, \&
  {Gotthelf}}]{2013ApJ...778..120H}
{Halpern}, J.~P., {Bogdanov}, S., \& {Gotthelf}, E.~V. 2013, \apj, 778, 120,
  \dodoi{10.1088/0004-637X/778/2/120}

\bibitem[{{Halpern} \& {Gotthelf}(2015)}]{2015ApJ...812...61H}
{Halpern}, J.~P., \& {Gotthelf}, E.~V. 2015, \apj, 812, 61,
  \dodoi{10.1088/0004-637X/812/1/61}

\bibitem[{{Hessels} {et~al.}(2007){Hessels}, {Stappers}, {Rutledge}, {Fox}, \&
  {Shevchuk}}]{2007A&A...476..331H}
{Hessels}, J.~W.~T., {Stappers}, B.~W., {Rutledge}, R.~E., {Fox}, D.~B., \&
  {Shevchuk}, A.~H. 2007, \aap, 476, 331, \dodoi{10.1051/0004-6361:20078330}

\bibitem[{{Lande} {et~al.}(2012){Lande}, {Ackermann}, {Allafort}, {Ballet},
  {Bechtol}, {Burnett}, {Cohen-Tanugi}, {Drlica-Wagner}, {Funk}, {Giordano},
  {Grondin}, {Kerr}, \& {Lemoine-Goumard}}]{2012ApJ...756....5L}
{Lande}, J., {Ackermann}, M., {Allafort}, A., {et~al.} 2012, \apj, 756, 5,
  \dodoi{10.1088/0004-637X/756/1/5}

\bibitem[{{Liu} {et~al.}(2022){Liu}, {Zeng}, {Xin}, \&
  {Zhang}}]{2022RvMPP...6...19L}
{Liu}, S., {Zeng}, H., {Xin}, Y., \& {Zhang}, Y. 2022, Reviews of Modern Plasma
  Physics, 6, 19, \dodoi{10.1007/s41614-022-00080-6}

\bibitem[{{Mereghetti} {et~al.}(2021){Mereghetti}, {Rigoselli}, {Taverna},
  {Baldeschi}, {Crestan}, {Turolla}, \& {Zane}}]{2021ApJ...922..253M}
{Mereghetti}, S., {Rigoselli}, M., {Taverna}, R., {et~al.} 2021, \apj, 922,
  253, \dodoi{10.3847/1538-4357/ac34f2}

\bibitem[{{Miller} \& {Bregman}(2013)}]{2013ApJ...770..118M}
{Miller}, M.~J., \& {Bregman}, J.~N. 2013, \apj, 770, 118,
  \dodoi{10.1088/0004-637X/770/2/118}

\bibitem[{{Ohira} {et~al.}(2010){Ohira}, {Murase}, \&
  {Yamazaki}}]{2010A&A...513A..17O}
{Ohira}, Y., {Murase}, K., \& {Yamazaki}, R. 2010, \aap, 513, A17,
  \dodoi{10.1051/0004-6361/200913495}

\bibitem[{{Porter} {et~al.}(2006){Porter}, {Moskalenko}, \&
  {Strong}}]{2006ApJ...648L..29P}
{Porter}, T.~A., {Moskalenko}, I.~V., \& {Strong}, A.~W. 2006, \apjl, 648, L29,
  \dodoi{10.1086/507770}

\bibitem[{{Porter} {et~al.}(2008){Porter}, {Moskalenko}, {Strong}, {Orlando},
  \& {Bouchet}}]{2008ApJ...682..400P}
{Porter}, T.~A., {Moskalenko}, I.~V., {Strong}, A.~W., {Orlando}, E., \&
  {Bouchet}, L. 2008, \apj, 682, 400, \dodoi{10.1086/589615}

\bibitem[{{Rengelink} {et~al.}(1997){Rengelink}, {Tang}, {de Bruyn}, {Miley},
  {Bremer}, {Roettgering}, \& {Bremer}}]{1997A&AS..124..259R}
{Rengelink}, R.~B., {Tang}, Y., {de Bruyn}, A.~G., {et~al.} 1997, \aaps, 124,
  259, \dodoi{10.1051/aas:1997358}

\bibitem[{{Rutledge} {et~al.}(2008){Rutledge}, {Fox}, \&
  {Shevchuk}}]{2008ApJ...672.1137R}
{Rutledge}, R.~E., {Fox}, D.~B., \& {Shevchuk}, A.~H. 2008, \apj, 672, 1137,
  \dodoi{10.1086/522667}

\bibitem[{{Shimwell} {et~al.}(2017){Shimwell}, {R{\"o}ttgering}, {Best},
  {Williams}, {Dijkema}, {de Gasperin}, {Hardcastle}, {Heald}, {Hoang},
  {Horneffer}, {Intema}, {Mahony}, {Mandal}, {Mechev}, {Morabito}, {Oonk},
  {Rafferty}, {Retana-Montenegro}, {Sabater}, {Tasse}, {van Weeren},
  {Br{\"u}ggen}, {Brunetti}, {Chy{\.z}y}, {Conway}, {Haverkorn}, {Jackson},
  {Jarvis}, {McKean}, {Miley}, {Morganti}, {White}, {Wise}, {van Bemmel},
  {Beck}, {Brienza}, {Bonafede}, {Calistro Rivera}, {Cassano}, {Clarke},
  {Cseh}, {Deller}, {Drabent}, {van Driel}, {Engels}, {Falcke}, {Ferrari},
  {Fr{\"o}hlich}, {Garrett}, {Harwood}, {Heesen}, {Hoeft}, {Horellou},
  {Israel}, {Kapi{\'n}ska}, {Kunert-Bajraszewska}, {McKay}, {Mohan},
  {Orr{\'u}}, {Pizzo}, {Prandoni}, {Schwarz}, {Shulevski}, {Sipior}, {Smith},
  {Sridhar}, {Steinmetz}, {Stroe}, {Varenius}, {van der Werf}, {Zensus}, \&
  {Zwart}}]{2017A&A...598A.104S}
{Shimwell}, T.~W., {R{\"o}ttgering}, H.~J.~A., {Best}, P.~N., {et~al.} 2017,
  \aap, 598, A104, \dodoi{10.1051/0004-6361/201629313}

\bibitem[{{Wood} {et~al.}(2017){Wood}, {Caputo}, {Charles}, {Di Mauro},
  {Magill}, {Perkins}, \& {Fermi-LAT Collaboration}}]{2017ICRC...35..824W}
{Wood}, M., {Caputo}, R., {Charles}, E., {et~al.} 2017, in International Cosmic
  Ray Conference, Vol. 301, 35th International Cosmic Ray Conference
  (ICRC2017), 824.
\newblock \doarXiv{1707.09551}

\bibitem[{{Xiang} \& {Jiang}(2021)}]{2021ApJ...908...22X}
{Xiang}, Y., \& {Jiang}, Z. 2021, \apj, 908, 22,
  \dodoi{10.3847/1538-4357/abd175}

\bibitem[{{Yang} {et~al.}(2014){Yang}, {Zhang}, {Yuan}, \&
  {Liu}}]{2014A&A...567A..23Y}
{Yang}, R.-z., {Zhang}, X., {Yuan}, Q., \& {Liu}, S. 2014, \aap, 567, A23,
  \dodoi{10.1051/0004-6361/201322737}

\bibitem[{{Yuan} {et~al.}(2014){Yuan}, {Huang}, {Liu}, \&
  {Zhang}}]{2014ApJ...785L..22Y}
{Yuan}, Q., {Huang}, X., {Liu}, S., \& {Zhang}, B. 2014, \apjl, 785, L22,
  \dodoi{10.1088/2041-8205/785/2/L22}

\bibitem[{{Zabalza}(2015)}]{2015ICRC...34..922Z}
{Zabalza}, V. 2015, in International Cosmic Ray Conference, Vol.~34, 34th
  International Cosmic Ray Conference (ICRC2015), 922.
\newblock \doarXiv{1509.03319}

\bibitem[{{Zane} {et~al.}(2011){Zane}, {Haberl}, {Israel}, {Pellizzoni},
  {Burgay}, {Mignani}, {Turolla}, {Possenti}, {Esposito}, {Champion},
  {Eatough}, {Barr}, \& {Kramer}}]{2011MNRAS.410.2428Z}
{Zane}, S., {Haberl}, F., {Israel}, G.~L., {et~al.} 2011, \mnras, 410, 2428,
  \dodoi{10.1111/j.1365-2966.2010.17619.x}

\bibitem[{{Zeng} {et~al.}(2019){Zeng}, {Xin}, \& {Liu}}]{2019ApJ...874...50Z}
{Zeng}, H., {Xin}, Y., \& {Liu}, S. 2019, \apj, 874, 50,
  \dodoi{10.3847/1538-4357/aaf392}

\bibitem[{{Zeng} {et~al.}(2021){Zeng}, {Xin}, {Zhang}, \&
  {Liu}}]{2021ApJ...910...78Z}
{Zeng}, H., {Xin}, Y., {Zhang}, S., \& {Liu}, S. 2021, \apj, 910, 78,
  \dodoi{10.3847/1538-4357/abe37e}

\end{thebibliography}
\bibliographystyle{aasjournal}

%% This command is needed to show the entire author+affiliation list when
%% the collaboration and author truncation commands are used.  It has to
%% go at the end of the manuscript.
%\allauthors

%% Include this line if you are using the \added, \replaced, \deleted
%% commands to see a summary list of all changes at the end of the article.
%\listofchanges
%\end{CJK}
\end{document}